\documentclass[12pt,oneside]{article}
\usepackage{amsfonts,amssymb,graphicx}
\def\be{\begin{equation}}
\def\ee{\end{equation}}
\def\bea{\begin{eqnarray}}
\def\eea{\end{eqnarray}}
\def\<{\langle}
\def\>{\rangle}
\def\~{\tilde}
\def\s{\sigma}

\def\t{\tau}

\def\ds{\displaystyle}
\begin{document}

\begin{center}{\sc
CONVEX REPLICA SIMMETRY BREAKING\\ FROM POSITIVITY
AND THERMODYNAMIC LIMIT}
\vskip 1truecm
\end{center}
\begin{center}{Pierluigi Contucci, Sandro Graffi}\\
\vskip 1truecm
{\small Dipartimento di Matematica} \\
    {\small Universit\`a di Bologna,
    40127 Bologna, Italy}\\
    {\small {e-mail:contucci@dm.unibo.it, graffi@dm.unibo.it}}
\end{center}
%
%
\begin{abstract}\noindent
Consider a correlated Gaussian random energy model built by successively adding 
one particle (spin) into the system and imposing the positivity of
the associated covariance matrix. We show that the validity of a recently isolated 
condition ensuring the existence of the thermodynamic limit forces
the covariance matrix to exhibit the Parisi replica symmetry breaking scheme
with a convexity condition on the matrix elements.
\end{abstract}
\newpage
The existence of the thermodynamic limit  has been
recently  proved for the Sherrington-Kirckpatrick (SK) model
\cite{1}, and more generally for any correlated Gaussian random
energy (CGREM) model including the Derrida REM and the Derrida-Gardner GREM
\cite{2}. In this letter we point out that the proof of
\cite{2} may shed some light on the origin of Parisi's algebraic ansatz for
the replica symmetry breaking (RSB) scheme which lies at the
basis his solution \cite{3} of the SK model.

Algebraically, Parisi's ansatz may be described as follows: start from
the one by one matrix $q(0)>0$. Take $q(1)$ as $0<q(0)<q(1)<1$ first replicate the
system
\be
R(q(0))= \left(\begin{array}{ll} q(0) & q(0)\\ q(0) &
q(0)\end{array}\right),
\ee
then break the symmetry
\be
\qquad BR(q(0))=\left(\begin{array}{ll}
q(1) & q(0)\\ q(0) & q(1)\end{array}\right).
\ee
If we iterate $n$ times the operation $BR$ we end up with the Parisi
RSB matrix of order $n$, $[BR]^n(q(0))=Q^{(n)}$. Example: for $p=3$ we
get the $8\times 8$ matrix
\be
Q^{(3)}=\left( \begin{array}{cccccccc}
    {q(3)} & {q(2)} & {q(1)} & {q(1)} & {q(0)} & {q(0)} & {q(0)} & {q(0)}
\cr {q(2)} & {q(3)} & {q(1)} & {q(1)} & {q(0)} & {q(0)} & {q(0)} & {q(0)}
\cr {q(1)} & {q(1)} & {q(3)} & {q(2)} & {q(0)} & {q(0)} & {q(0)} & {q(0)}
\cr {q(1)} & {q(1)} & {q(2)} & {q(3)} & {q(0)} & {q(0)} & {q(0)} & {q(0)}
\cr {q(0)} & {q(0)} & {q(0)} & {q(0)} & {q(3)} & {q(2)} & {q(2)} & {q(2)}
\cr {q(0)} & {q(0)} & {q(0)} & {q(0)} & {q(2)} & {q(3)} & {q(2)} & {q(2)}
\cr {q(0)} & {q(0)} & {q(0)} & {q(0)} & {q(2)} & {q(2)} & {q(3)} & {q(2)}
\cr {q(0)} & {q(0)} & {q(0)} & {q(0)} & {q(2)} & {q(2)} & {q(2)} & {q(3)}
\end{array}%
\right),
\ee
As is well known, the RSB method consists in solving the saddle point
equations for the SK
free energy after  $n$ replicas assuming that the saddle point matrix
has the form
$Q^{(n)}$; the maximum is to be found  among the functions on
$[0,1]$ taking the values $q(k)$ at the prescribed points $m_k$ at
 step $k$ of RSB.  Performing first the limit $k\to\infty$ and then
the limit
$n\to 0$ one gets:
\be
\lim_{n\to 0}\lim_{k\to\infty}\frac1{n}\sum_{a,b=1}^nQ^{k}_{a,b}=
-\int_0^1q(x)\,dx
\label{gp}
\ee
$q(x)$ is continuous on $[0,1]$ and is the (Parisi) order parameter of the
local square magnetization because its inverse function
$x(q)$ is such that
$\ds  x(q)=\int_{0}^q P(s)\,ds$ where
$P(q)$ is the overlap probability distribution between pure
magnetization states. Hence $q(x(q))=q$ and the  elements
$q(1),\ldots,q(k)$ are overlaps.

Even though it has been recently proved that the Parisi free energy is at
least a lower bound of the SK free energy\cite{4}, the RSB is still far
from a mathematical understanding. A major puzzle is still represented
by the origin of the algebraic structure of the Parisi ansatz, which lies at
the basis also of the Derrida-Gardner GREM\cite{5}.

In this letter we point out that the algebraic Parisi ansatz and,
equivalently, the Derrida-Gardner GREM are generated by a precise
prescription to reach monotonically the thermodynamic limit with a family of
correlated gaussian random energy models (CGREM).

More precisely, we will build by recurrence over $N$, increasing the size
by $1$ at each step (equivalently,  adding  a spin) a family $E_\sigma(N)$
of CGREM for which we require:
\par\noindent
1.  $E_\sigma(N)$
($N$-spins) has to be a subsystem of $E_\sigma(N+1)$ ($N+1$-spins);
\par\noindent
2. The sequence $E_\sigma(N)$  fulfills the conditions ensuring the
existence of the thermodynamic limit.
\par\noindent
The two previous conditions imply that the matrix sequence $C_N$ fulfills the RSB
scheme, in the
sense that for each $N$ the correlation matrix $C_N$  fulfills the Parisi
ansatz at step $N$. In particular $E_\sigma(N)$ can be identified with the
Derrida-Gardner GREM with a convex growth scheme.

\noindent
For $N=0$ the system is the {\it vacuum} (no particles) with random energy
\be
E_0 = \xi_0(0)
\ee
 $\xi_0$ is a centered Gaussian variable with variance $c(0)$, which
we can consider as a one-by-one matrix. We include the system of
$0$ spins into a new one-spin system (for short: we add a spin) assuming
at first that the one-spin Hamiltonian doesn't depend on the added spin
\be
\tilde E_1(\s_1) = \xi_0(0) \; ;
\ee
the covariance matrix of this process is
\be
\tilde C(1)=\left( \begin{array}{cc} {c(0)} & {c(0)} \cr {c(0)} & {c(0)} \cr
\end{array}
\right) \; ;
\ee
This construction was called {\it lifting} in \cite{2}.

The one-spin system is a CGREM of size one only if
its covariance matrix is a (non-degenerate) positive definite matrix. The
elimination of the
degenarcy requires the introduction in the Hamiltonian of a new Gaussian variable
$\xi_1$
parametrizing the dependence on the added spin:
\be
E_1(\s_1) = \xi_0(1) + \xi_1(1)\s_1 \; ;
\label{1}
\ee
the covariance matrix element is
\be
Av(E_1(\s_1)E_1(\t_1)) = Av(\xi_0(1)^2) + Av(\xi_1(1)^2)\s_1\t_1 \; ;
\ee
Since
\be
\s\t = 2\delta_{\s,\t} -1
\ee
we have
\be
Av(E_1(\s_1)E_1(\t_1)) = Av(\xi_0(1)^2) - Av(\xi_1(1)^2) +
2Av(\xi_1(1)^2)\delta_{\s_1,\s_2} \;
\ee
which we will write as
\be
Av(E_1(\s_1)E_1(\t_1)) = a_0(1) + a_1(1)\delta_{\s_1,\s_2} \; .
\ee
Defining $c(l) = \sum_{k=0}^{l}a_k$ we have

\be
C(1)=\left( \begin{array}{cc} {c(1)} & {c(0)} \cr {c(0)} & {c(1)} \cr
\end{array}%
\right),
\ee
with $c(1)>c(0)$

To  iterate the procedure, let us first describe the second step, i.e. the
addition of a second spin to build a 2-spin system. As before, we first
assume independence on  the newly added spin
variable. The covariance matrix turns then out to be
(with the lexicographic order of the spin configurations)
\be
\tilde C(2)=\left( \begin{array}{cccc} {c(1)} & {c(1)} & {c(0)} & {c(0)} \cr
{c(1)} &
{c(1)} & {c(0)} & {c(0)}
\cr {c(0)} & {c(0)} & {c(1)} & {c(1)}
\cr {c(0)} & {c(0)} & {c(1)} & {c(1)} \cr
\end{array}%
\right) \; .
\ee
Again, this matrix is the covariance of a CGREM process only if it positive
definite and non-degenerate. As above, this requires the dependence on the second
spin
variable.  Among the possible ways to parametrize this dependence we
choose the {\it minimal} one, namely the preceding one which only modifies
the subprincipal diagonals:
\be
\bar C(2)=\left( \begin{array}{cccc} {c(1)} & {\bar q_2} & {c(0)} & {c(0)}
\cr {\bar q_2} &
{c(1)} & {c(0)} & {c(0)}
\cr {c(0)} & {c(0)} & {c(1)} & {\bar q_2}
\cr {c(0)} & {c(0)} & {\bar q_2} & {c(1)} \cr
\end{array}%
\right),
\ee
with $c(0)<\bar q_2<c(1)$. We relabel the elements $c(1)\to
c(2)$ and
$\bar q_2\to c(1)$ obtaining the final 2-spin covariance matrix
\be
C(2)=\left( \begin{array}{cccc} {c(2)} & {c(1}) & {c(0)} & {c(0)} \cr {c(1)}
&
{c(2)} & {c(0)} & {c(0)}
\cr {c(0)} & {c(0)} & {c(2)} & {c(0)}
\cr {c(0)} & {c(0)} & {c(1)} & {c(2)} \cr
\end{array}%
\right),
\ee
with $c(2)>c(1)>c(0)>0$. This last condition implies that the matrix is
positive
definite because the 4 principal minors are
\be
\Delta_1=c(2)>0\; ,
\ee
\be
\Delta_2=c(2)^2-c(1)^2>0 \; ,
\ee
\be
\Delta_3=[(c(2)-c(1)][c(2)(c(1)+c(2))-2c(0)^2]>0\;
\ee
and
\be
\Delta_4=[c(2) - c(1)]^2[((c(1)+c(2))^2 -4 c(0)^2]>0
\ee
Corrispondigly with the spin representation we would have
\be
E_2(\s_1,\s_2) = \xi_0(2) + \xi_1(2)\s_1 + \xi_2(2)\s_2 + \xi_{1,2}\s_1\s_2 \; ,
\ee
and
\bea\nonumber
Av(E_2(\s_1,\s_2)E_2(\t_1,\t_2)) &=& Av(\xi_0(2)^2) - Av(\xi_1(2)^2) -
Av(\xi_2(2)^2) +
Av(\xi_{1,2}(2)^2)\\ \nonumber
&+& \delta_{\s_1,\t_1}[2Av(\xi_1(2)^2)-2Av(\xi_{1,2}(2)^2)] \\ \nonumber
&+& \delta_{\s_2,\t_2}[2Av(\xi_2(2)^2)-2Av(\xi_{1,2}(2)^2)] \\ \nonumber
&+& \delta_{\s_1,\t_1}\delta_{\s_2,\t_2} 4Av(\xi_{1,2}(2)^2)
\; ;
\eea
and since we choose
\be
Av(\xi_2(2)^2)=2Av(\xi_{1,2}(2)^2)
\ee
the covariance matrix exibit the RSB scheme:
\be
Av(E_1(\s_1,\s_2)E_1(\t_1\t_2)) = a_0(2) + a_1(2)\delta_{\s_1,\s_2} +
a_2(2)\delta_{\s_1,\t_1}\delta_{\s_2,\t_2} \; .
\ee
In general we will assume that our construction
is done adding at each step the $N$-th spin variable and the newly added interaction
terms are indipendent realization of the same Gaussian distribution or, in other
terms,
the distribution of the Gaussian random variables depends only on the last index.
The general scheme is the described by a correlated Gaussian process
\be
E_N(\s) = \xi_0(N) + \sum_{1\le i\le N}\xi_i(N)\s_i + \sum_{1\le i<j \le
N}\xi_{i,j}(N)\s_i\s_j +\dots
+\xi_{1,2,\dots,N}\s_1\cdots\s_N\
\ee
in which the distribution of each Gaussian variable depends only on the
last index; its covariance matrix:
\be
Av(E_N(\s)E_N(\t)) = \sum_{k=0}^{N}a_{k}(N)\prod_{i=0}^{k}\delta_{\s_i,\t_i} \; ,
\label{gc}
\ee
turns out ot fulfill the RSB structure
An important observation is that we may, within the positivity
conditions, chose a subset of c's that fulfill the conditions for the existence
of the thermodynamical limit as shown in \cite{2}. Here we observe that those
conditions translate into a convexity
property for the matrix elements. In fact
introducing the
two complementary liftings (labelled as left $l$ and right $r$) of the one-spin system:
\be
C_l(2)=\left( \begin{array}{cccc} {c(2)} & {q(2}) & {c(0)} & {c(0)} \cr
{c(2)} &
{c(2)} & {c(0)} & {c(0)}
\cr {c(0)} & {c(0)} & {c(2)} & {c(2)}
\cr {c(0)} & {c(0)} & {c(2)} & {c(2)} \cr
\end{array}%
\right),
\ee
and
\be
C_r(2)=\left( \begin{array}{cccc} {c(2)} & {q(0}) & {c(2)} & {c(0)} \cr
{c(0)} &
{c(2)} & {c(0)} & {c(2)}
\cr {c(0)} & {c(2)} & {c(2)} & {c(0)}
\cr {c(2)} & {c(0)} & {c(0)} & {c(2)} \cr
\end{array}%
\right),
\ee
the conditions developed in \cite{2} are, for the symmetric sub-division of the
system into 2 subsystems of the same size,
\be
C(2)\le \frac{1}{2} [C_l(2)+C_r(2)]
\ee
(component-wise) which imposes
\be
c(1)\le \frac{1}{2}[c(0)+c(2)] \; .
\ee
For general $N$ the conditions in \cite{2} turn out to be a set of $2^N$ relations between the matrix
elements:
\be
Q(N)\le \frac{1}{2} [C_l(N)+C_r(N)]
\ee
With respect to the size $N-1$ the only new relationships required only
involve the quantity
$c(N), c(N-1), c(N-2)$. Indeed at step N the only new inequalities (with
respect to those at step
$N-1$) are those coming from the four by four principal minors of the three
matrices. After the
operation  of re-enumeration these inequalities become (element wise)
\be
Q^{(4)}(N)\le \frac{1}{2} [C_l^{(4)}(N)+C_r^{(4)}(N)]
\ee
with
\be
Q^{(4)}(N)=\left( \begin{array}{cccc} {c(N)} & {c(N-1)} & {c(N-2)} &
{c(N-2)} \cr {c(N-1)} &
{c(N)} & {c(N-2)} & {c(N-2)}
\cr {c(N-2)} & {c(N-2)} & {c(N)} & {c(N-1)}
\cr {c(N-2)} & {c(N-2)} & {c(N-1)} & {c(N)} \cr
\end{array}%
\right),
\ee
\be
C_l^{(4)}(N)=\left( \begin{array}{cccc} {c(N)} & {q(N}) & {c(N-2)} &
{c(N-2)} \cr {c(N)} &
{c(N)} & {c(N-2)} & {c(N-2)}
\cr {c(N-2)} & {c(N-2)} & {c(N)} & {c(N)}
\cr {c(N-2)} & {c(N-2)} & {c(N)} & {c(N)} \cr
\end{array}%
\right),
\ee
and
\be
C_r^{(4)}(N)=\left( \begin{array}{cccc} {c(N)} & {q(N-2}) & {c(N)} &
{c(N-2)} \cr {c(N-2)} &
{c(N)} & {c(N-2)} & {c(N)}
\cr {c(N)} & {c(N-2)} & {c(N)} & {c(N-2)}
\cr {c(N-2)} & {c(N)} & {c(N-2)} & {c(N)} \cr
\end{array}%
\right),
\ee
from which we have
\be
c(N-1)\le \frac{1}{2}[c(N)+c(N-2)] \; .
\ee
Equivalently in the case of general partition of the $N$-particle 
systems into two subsystems of size $N_1$ and $N_2$ the existence of the 
thermodynamic limit will be assured by
\be
c(N-1)\le \frac{N_1}{N}c(N)+\frac{N_2}{N}c(N-2) \; ,
\ee
so that any positive increasing sequence $c(i)$, $i=1,...,N$ with
the convexity property
\be
c(i)\le \alpha_i c(i+1)+(1-\alpha_i)c(i-1) \; ,
\ee
for suitable chosen $0<\alpha_i<1$ provides a CGREM with RSB structure whose
thermodynamic limit is reached monotonically.\\
Remark: introducing \cite{5} the ultrametric overlap 
\be
d_N(\s,\t)\; = \; \frac{1}{N}\min(i|\s_i\neq \t_i)
\label{um}
\ee
one can see that the covariance matrix (\ref{gc}) is a growing
function of $d_N$ for every choice of the $a$'s. Obviously at zero temperature
the overlap (\ref{um}) coincides with the standard GREM overlap 
among magnetization states.\\\\
{\bf Acknowledgements}\\
We want to thank Francesco Guerra for many enlightening discussions.

\end{document}